\documentclass[english]{llncs}

\usepackage[T1]{fontenc}

\usepackage{times}

\usepackage[normalem]{ulem} 
\usepackage{xcolor}

\usepackage{ifthen} 
\usepackage{amssymb}
\newboolean{showcomments}
\setboolean{showcomments}{true} 
\ifthenelse{\boolean{showcomments}}
  {\newcommand{\nb}[2]{
    \fcolorbox{gray}{yellow}{\bfseries\sffamily\scriptsize#1}
    {$\blacktriangleright$#2$\blacktriangleleft$}
   } 
   
  }
  {\newcommand{\nb}[2]{}
   
  }

\usepackage{url} 
\usepackage{hyperref}
\usepackage{verbatim}

\usepackage{amsmath}
\usepackage{amssymb}	

\usepackage{epsfig}
\usepackage{subfigure}
\usepackage{array}			
\usepackage{multirow}
\usepackage{rotating}

\newtheorem{defn}{Definition}

\usepackage{algorithm}
\usepackage{algorithmic}

\newcommand{\ie}{\emph{i.e.,~}}							
\newcommand{\etal}{~\emph{et al.}}					
\newcommand{\Fig}[1]{Fig.~\ref{#1}}  			
\newcommand{\Table}[1]{Table~\ref{#1}}	    
\newcommand{\Model}[1]{\textsf{\small{#1}}} 
\newcommand{\Code}[1]{\texttt{\small{#1}}}	

\newcommand{\ATOM}{AToMPM}

\newcommand{\Virtual}{\Code{\guillemotleft virtual\guillemotright}}
\newcommand{\AVirtual}{\Code{\guillemotleft abstract, virtual\guillemotright}}
\newcommand{\SVirtual}{\Code{\guillemotleft standard, virtual\guillemotright}}

\newcommand{\Abstract}{\Code{\guillemotleft abstract\guillemotright}}
\newcommand{\Standard}{\Code{\guillemotleft standard\guillemotright}}
\newcommand{\Locked}{\Code{\guillemotleft locked\guillemotright}}

\begin{document}

\frontmatter

\title{Towards Controlling Refinements of Statecharts}

\author{Conner Hansen\inst{1} \and Eugene Syriani\inst{1} \and Levi Lucio\inst{2}}
\institute{University of Alabama, Tuscaloosa AL, U.S.A.
          \and McGill University, Montreal, Canada\\
	\email{chansen@crimson.ua.edu,esyriani@cs.ua.edu,levi@cs.mcgill.ca}
}

\maketitle

\begin{abstract}
In incremental development strategies, modelers frequently refine Statecharts models to satisfy requirements and changes. Although several solutions exist to the problem of Statecharts refinement, they provide such levels of freedom that a statechart cannot make assumptions or guarantees about its future structure. In this paper, we propose a set of bounding rules to limit the allowable Statecharts refinement operations such that certain assumptions will hold.
\end{abstract}

\section{Introduction} \label{sec:introduction}

The Statecharts formalism is considered a defacto standard for modeling reactive systems~\cite{Harel1987}.
As in any modern software development, statecharts are developed incrementally.
Each incremental modification, within bounds of the rules defined in this paper, can be viewed as a \emph{refinement} of the previous version.
Therefore from a reusability and maintenance perspective, it is crucial to control how modelers can refine statechart models.
UML~\cite{OMG2001} and Rhapsody~\cite{rhapsody} have addressed the topic of Statecharts refinement, but in a totally different context of class inheritance which we consider as specialization.

The goal of our work is to provide a built-in mechanism to assist modelers with incrementally designing statecharts.
The refinement approach we propose abides to basic software engineering design principles, such as the \emph{open-closed principle} stating that the model should be ``open for extension, but closed for modification''.
We therefore explore the concept of Statecharts refinement as an analog for extension within the Object-Oriented (OO) paradigm.
Within the OO world, the class-subclass structure that results from the associated extension rules comes with a certain set of expectations.
For instance, if there are unimplemented functions in the superclass, then a subclass must implement them if it is to be usable.
In situations where the instance of the superclass is expected, any subclass (or descendent) instance must be able to stand in for that expectation as stipulated by the Liskov \emph{substitutability principle}~\cite{Liskov1987}.

Our main motivation is to provide a set of rules that govern Statecharts refinement with the intention of preserving the external behavior of the original statechart.
Focusing on this preservation has the potential to increase the predictability of the to-be-refined statecharts, as well as respecting the expectations of the original statechart.
Therefore we consider the refinement process to be purely additive: \ie no removal of elements from the original is allowed.
Refinement is not meant as a replacement for editing which follows an entirely distinct process that should be performed on the original statechart rather than on the refined.
These rules must also be focused on providing realistic and usable boundaries, though empirical studies will have to be done at a later time.


\section{Refinement Relation}\label{sec:rationale}

As defined in~\cite{Harel1987}, a statechart model is defined by $SC = \left\langle S, T, E, V, P, g, en, ex, in \right\rangle$.
It consists of a set of states $S$, transitions $T$, events $E$, and variables $V$.
$P$ is the set of predicates built on type values, type variables, and boolean negation and conjunction.
There are two kinds of events: external events are provided as stimuli from the context outside the statechart and internal events are signals generated by the statechart.
States can be basic, pseudo, OR or AND.
$in: S \rightarrow \wp(S)$ denotes the containment relationship which must be non-empty for OR-states and AND-states.
Non-pseudo states can define entry and exit actions $en,ex: S \rightarrow V$ over the set of variables.
Transitions $T: S \times E \times \wp(E) \rightarrow \wp(S)$ define the evolution between states.
They can be guarded $g: T \rightarrow P$ by a predicate over variables, triggered by events, and output event signals.
As defined in~\cite{Harel1996}, a configuration is a legal snapshot of the statechart before or after a microstep.

Statechart refinement is a restricted form of modification of the original model such that design decisions are preserved in the refined model.
This is ensured by the refinement conditions in definition~\ref{def:refinement}.
To define the refinement relation between two statechart models, we consider their flattened versions using the Statemate semantics~\cite{Harel1987a,Wasowski2004}.
We denote the flattened version of $SC$ by $flat(SC) = \langle flat(S), flat(T), E, V,$ $P, g, en, ex \rangle$, where $flat(S)$ consists of basic states only and $flat(T): flat(S) \times E \times \wp(E) \rightarrow flat(S)$.
In the following, we denote $SC_o$ and $SC_r$ respectively the original and refined flattened statecharts.

\begin{defn}[Refinement Relation]
\label{def:refinement}
  The refinement $R : SC_o \rightarrow SC_r$ includes the relation
  $R \subseteq S_o \times S_r \cup S_o \times (S_r \cup T_r) \cup T_o \times T_r \cup T_o \times (S_r \cup T_r)$, such that the following refinement conditions are satisfied.
  We denote the set of all refinements for all statecharts as $\Re$.
\vspace*{-.5\baselineskip}%
\begin{description}
	\item[Inverse Surjection]
  $\forall st_r \in S_r \cup T_r, \exists! st_o \in S_o \cup T_o: R^{-1}(st_r) = st_o$
  \item[Event and Variable Inclusion]
  $E_o \subseteq E_r \wedge V_o \subseteq V_r$
  \item[Guard Inclusion]
  $\forall t_r \in T_r, \exists t_o \in T_o : g_r(t_r) = g_o(t_o) \wedge p, \mbox{for } p \in P_r$
  \item[Structural Inclusion]:\\
    $\mbox{Let } LTS\left(\left\langle S,T,E,V,P,g,en,ex \right\rangle\right) = \left\langle flat(S),flat(T),E,P \right\rangle.$\\
    $\mbox{If } s_o\stackrel{e[p_o]/x}{\longrightarrow}s_o' \in LTS(SC_o), \mbox{ then }$
    $\forall (s_o,s_r) \in R \wedge s_r \in S_r:$ \\
    $s_r\stackrel{e[p_r]}{\longrightarrow}s_r''\stackrel{*}{\longrightarrow}s_r'''\stackrel{[p_r]/x}{\longrightarrow}s_r' \in LTS(SC_r)^{*}
    \mbox{~~or~~} s_r\stackrel{e[p_r]/x}{\longrightarrow}s_r' \in LTS(SC_r)$ \\
    $\mbox{where } s_o, s'_o \in S_o, s_r, s'_r, s''_r, s'''_r \in S_r, p_o,p_r \in P$
\end{description}
\end{defn}

The first condition ensures that any state or transition in $SC_r$ is mapped to exactly one state or transition in $SC_o$.
Therefore the inverse refinement $R^{-1}$ is a surjection.
The second condition ensures that all original events and variables must be preserved at minimum.
The third condition guarantees that at least the original guard is preserved.
A consequence of the fourth condition is that for any states $s_o, s_o' \in S_o$, if there is a path from $s_o$ to $s_o'$ in $SC_o$, then there must be a path from $R(s_o)$ to $R(s_o')$ in $SC_r$.
Note that $LTS(SC)^{*}$ is the transitive closure of the paths in $LTS(SC)$.
We do not consider intermediate guards for the structural inclusion to allow design choices in the implementation, such as conjunction, modification, or preservation of guards.
Although this is abstracted in our formalization, this may lead to certain runs of execution not being achievable.

When the above conditions are met, we say that $SC_r$ preserves the \emph{external behavior} of $SC_o$.
An example of this preservation, is that if the current configuration of $SC_r$ is in a state \Model{A} and it receives an event \Model{b}, if the original behavior would end the microstep in state \Model{B} (and ignores all other events), then the refined behavior must also end in state \Model{B} even if the sequence of events received has new, intermediate events.
A substate of \Model{B} is acceptable as well if \Model{B} has also been refined.
Consequently, the Statechart refinement allows for the following possible modifications: new events can be interjected between events from the original (\ie new microsteps are possible), new states can be added, and new transitions can be added.
However, these additions are subject to specific constraint rules in order to reflect the intention of incremental modeling.

A notable property of $R$ is that it is reflexive: a statechart can be refined into itself (or a copy of itself), in which case $SC_r = SC_o$.
This allows one to refine only parts of a statechart while keeping others intact as in the original.
$R$ is also transitive: if $SC_2$ is a refinement of $SC_1$ and $SC_3$ is a refinement of a $SC_2$, then $SC_3$ is a refinement of $SC_1$.
This enables modelers to define multi-step refinements and therefore allows for incremental modeling of Statecharts.

\vspace{-.5\baselineskip}
\section{Statecharts Refinement}\label{sec:refine}
\vspace{-.5\baselineskip}

In the previous section, we laid out the requirements for refinement rules.
In this section, we extend Statecharts with refinement annotations and define the rules for producing a well-formed refined statechart.
These rules are intended to be statically validated, though some requirements are either difficult to or cannot be validated statically.

\vspace{-.5\baselineskip}
\subsection{Extending Statecharts with Refinement Modifiers}\label{sec:Modifiers}
\vspace{-.25\baselineskip}

In Statemate, Rhapsody, or UML State Machines formalisms, there is currently no way to identify a Statecharts element
as needing further information added before it is ready to be utilized.
This requires those who design and know the system to track what components need further details and which are ready for operation.
We propose four modifiers, three of which can be applied as stereotypes to states, pseudo-states, and transitions.
The fourth modifier can only be applied automatically by the tool as a stereotype to AND- or OR-states and must be combined with another modifier.
Note that the modifiers are annotations to the statechart and, without altering its behavior, simply indicate elements that are possible or required to be refined to the modeler.
The statechart is still executable, since it must still be well-formed regardless of the presence of modifiers.

\noindent\textbf{Abstract}
An \Abstract{} modifier identifies a state (basic, OR or AND) or transition that requires implementation details in order to be usable.
As with an abstract class within an OO language, this modifier is intended to convey clearly to the user of the statechart that the particular element in question must be refined before it can be accessed and run.
As with abstract classes in OO languages, abstract states and transitions can themselves be fully implemented with no restriction.
Any abstract element within a statechart must be refined into a non-abstract element when refining the statechart in order for it to be considered fully implemented.

\noindent\textbf{Standard}
A \Standard{} modifier identifies a Statecharts element that can be considered implemented.
This modifier allows for refinement, though any details present within the original elements must be preserved.

\noindent\textbf{Locked}
A \Locked{} modifier identifies a Statecharts element that is considered implemented and non-refinable.
This modifier is analogous to the \emph{final}~\cite{Gosling2005} or \emph{sealed}~\cite{CSharp2001} modifiers in modern OO languages.
Therefore, when refining a statechart with locked elements, those elements will be copied as is in the refined model.

\noindent\textbf{Virtual}
A \Virtual{} modifier is a pseudo-modifier that is reserved by the tool that indicates a state that was automatically generated and that may require additional details outside the bounds of what standard refinement allows for.
This modifier cannot be applied to transitions, as transitions cannot act as containers for other elements.
This modifier is added to a basic state that refines into an AND- or OR-state, to any newly created orthogonal components inside the refinement of an AND-state or inside the refinement of an OR-state to an AND-state.
New transitions can be added between states that already exist within the marked state, so long as the source and target of the transition are within the \Virtual{} state.
New states can also be added, so long as the resulting addition does not create an invalid statechart.
This is provided in order to cover those situations where significant structural freedoms must be granted to (potentially many) future refinements.
This does not break the external behavior of the statechart, as these freedoms are only granted within the \Virtual{} state.

The \Virtual{} modifier is applied orthogonally alongside \Abstract{} and \linebreak \Standard{}: this gives rise to \AVirtual{} and \SVirtual{} modifiers, respectively.
Their meaning is exactly the same as the respective modifier, only now with the additional freedoms that \Virtual{} grants.
Since all refinements must preserve the \Locked{} modifier, unless explicitly stated otherwise, any substate that overrides the original \Locked{} modifier will also override the \Virtual{} modifier.
This allows for \Abstract{} states and transitions to exist within an otherwise \Locked{} state, while also preventing those states from granting themselves the \Virtual{} freedoms.

\vspace{-\baselineskip}%
\begin{table}
  \centering
  {\scriptsize
	\begin{tabular}{p{4mm}lccccc}
		\hline\hline
		&&&&\\ [-1.75ex]
		&& \multicolumn{4}{c}{\textbf{Original}} \\ 
		&& \textbf{Abstract} & \textbf{Abstract-Virtual} & \textbf{Standard} & \textbf{Standard-Virtual} & \textbf{Locked} \\ 
		\hline
		&&&&\\ [-1.75ex]
		\multirow{5}{*}{\begin{sideways}\textbf{Refines to}\end{sideways}} & \textbf{Abstract} & Yes & Yes & No & No & No \\ 
		& \textbf{Abstract-Virtual}	& By tool & Yes & No & No & No \\ 
		& \textbf{Standard} & Yes & Yes & Yes & Yes & No \\ 
		& \textbf{Standard-Virtual} & By tool & Yes & By tool & Yes & No \\ 
		& \textbf{Locked} & Yes & Yes & Yes & Yes & Yes \\ 
		\hline\\[-1ex]
	\end{tabular}}
	\caption{Valid refinements of modifiers.}
	\label{table:Hierarchy}
  \vspace{-2\baselineskip}%
\end{table}
\noindent\textbf{Modifier Application and Propagation}
All Statecharts elements have a modifier applied to them, though this application need not be explicit.
We assume that the default Statecharts modifier is \Standard{} if no modifiers are explicitly applied.
The modifier of a composite element (OR- or AND-state) is implicitly propagated to its contained elements, unless an inner element is assigned another modifier explicitly.
This is intended to reduce the number of modifiers in the diagrams.
Additionally, we view this as a more sensible development approach by requiring exceptions to be explicitly marked.
This way, a specific subelement of the hierarchy of a \Locked{} composite element can be explicitly made \Abstract{} or \Standard{}, while leaving the locked behavior of the rest of the composite element and its inner elements intact. 
These design decisions were made so that any tool implementing the refinement method we define will be able to utilize statechart models that were not defined with these modifiers in mind.

\Table{table:Hierarchy} illustrates the permissible refinements of modifiers.
The \Virtual{} modifier can only be applied to refinements of an already \Virtual{} state, newly created states, orthogonal components, or the refinement of a basic state to a OR-state.
Otherwise, no state can be refined to a less certain modifier as solutions should only become more concrete through refinement. Any substates of an AND- or OR-state that do not have an explicitly marked modifier will inherit the modifier of its super state. 

\vspace{-\baselineskip}
\subsection{Refinement Rules}\label{sec:Rules}

In what follows, we describe the rules governing the refinement of Statecharts.

\begin{figure}
	\centering
  \vspace{-1.5\baselineskip}
	\epsfig{file=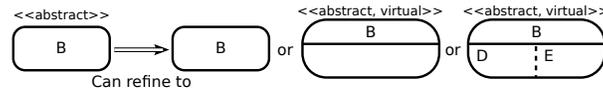, width=.65\linewidth}
	\caption{The generic basic state refinement.}
	\label{fig:basicState}
  \vspace{-1.5\baselineskip}
\end{figure}
\noindent\textbf{R1: Basic State}\label{sec:BasicState}
In \Fig{fig:basicState}, we show an overview of the three possibilities that a basic state can be refined into: a basic state, an OR-state, or an AND-state.
In the case of refinement into an OR- or AND-state, all newly created regions are given the \Virtual{} modifier by default.
The modifier of the resulting state must follow the rules in \Table{table:Hierarchy}.

Any defined actions of the source model are copied over to the refined state by default.
Each of these actions can be modified as long as it satisfies the principle of closed behavior from OO software.
We cannot verify the weakening of the pre-conditions or the strengthening of the post-conditions of an action, however it should be considered well-formed if these conditions are met. 
Additionally, all types of states (pseudo-states included) must have their names and in/output transitions preserved across a refinement.

\begin{figure}
	\centering
  \vspace{-\baselineskip}
	\epsfig{file=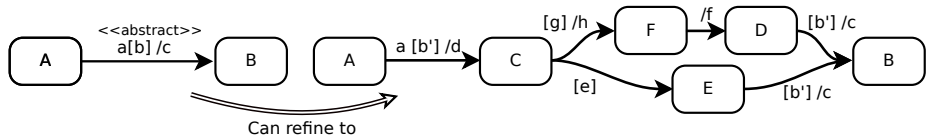, width=\linewidth}
	\caption{The generic transition state refinement.}
	\label{fig:transition}
  \vspace{-1.5\baselineskip}
\end{figure}
\noindent\textbf{R2: Transitions}\label{sec:Transition}
In \Fig{fig:transition} we demonstrate a possible transition refinement.
Any number of states can be inserted in between the transition's source and target states, however the execution flow through these states must pass from the source to the target as it would have in the original statechart.
The only exception to this case is when a composite element containing the source state of this transition (if such a container exists) has an outgoing transition of its own triggered, thereby prematurely ending the execution flow.
This exception is practical, because otherwise this behavior would significantly hinder the implementor when using Statemate's \emph{outer-first} semantics~\cite{Harel1997}. 

\noindent\textbf{R2.1: Events}\label{sec:TransitionEvents}
The original event must be preserved on the first outgoing transition.
In \Fig{fig:transition}, the original event, \emph{a}, is preserved after the refinement operation. Any intermediate transitions must be executed as microsteps, leading to an arrival at the original target state at the end of the processing of that particular trigger (plus the additional intermediate microsteps inserted by the refinement).

\noindent\textbf{R2.2: Guard}\label{sec:TransitionGuards}
The original guard of the transition must be preserved at a minimum, though new conjunctive statements are allowed.
This preserves, at a maximum, the original guard's state space.
For example, in \Fig{fig:transition}, guard $b$ is refined into guard $b'$, where either $b'=b$ or $b'=b \wedge p$ for some predicate $p \in P$.
Disjunctions are not allowed since that can expand the state space without limit effectively allowing for the removal of the guard.

\noindent\textbf{R2.3: Broadcast}\label{sec:TransitionBroadcast}
Broadcasts can be added to transitions created during refinement, excepting transitions connected to the original transition's target state (such as the new transitions from \Model{D} to \Model{B} and from \Model{E} to \Model{B} in \Fig{fig:transition}).
For those transitions, the original broadcast---or lack of a broadcast---must be preserved so that the behavior expected by the original target state remains preserved.
In \Fig{fig:transition}, the original broadcast $c$ is preserved along the final transitions of the refinement.
This is to preserve any expectation of the original statechart that a broadcast will be sent out just before the arrival at the destination state.
Note that, as stated in R2, events \Code{h} or \Code{f} may interrupt the execution flow to state \Code{B}.

\noindent\textbf{R2.4: Mutually Exclusive Internal Guards}\label{sec:TransitionExclusive}
Any guards added to transitions that are not connected to the original source or target states must have a mutually exclusive alternative path so that the execution flow is guaranteed.
That is, the original reachability is preserved in compliance with Definition~\ref{def:refinement}.
In \Fig{fig:transition}, there are two potential paths that could be taken from the added state C.
The transitions have guards $g$ and $e$, respectively.
Therefore, the relation $g=\neg e$ must hold so that there will always be a transition that can be activated. For our implementation, we can only verify the literal negation of the guard. Future implementations would need to be more logically flexible.

\begin{figure}
  \vspace{-1.5\baselineskip}
	\centering
  \epsfig{file=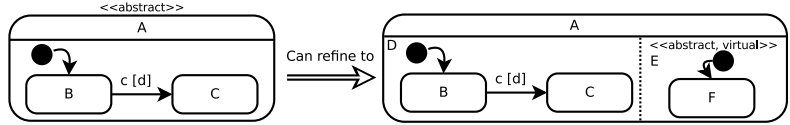, width=.85\linewidth}
  \caption{Refinement of an OR-state into an AND-state.}
  \label{fig:compositeState}
  \vspace{-1.5\baselineskip}
\end{figure}
\noindent\textbf{R3: OR-State}\label{sec:CompositeState}
An OR-state may be refined into an AND-state with two orthogonal components where one orthogonal component contains all of the original inner elements and the other only holds a region marked as \AVirtual{}.
Otherwise, the refinement is propagated to the inner elements of the OR-state as in \Fig{fig:compositeState}.
The dashed boundary around the internal states is meant to convey that all of the elements encompassed each have the same modifier applied.

\begin{figure}
	\centering
	\epsfig{file=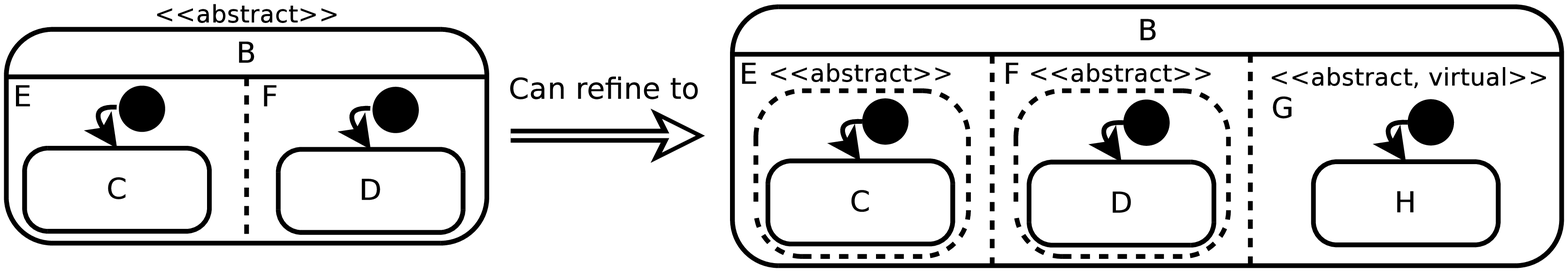, width=.85\linewidth}
	\caption{The generic AND-state refinement.}
	\label{fig:orthoComp}
  \vspace{-1.5\baselineskip}
\end{figure}
\noindent\textbf{R4: AND-state}\label{sec:OrthogonalComp}
In \Fig{fig:orthoComp} we demonstrate the refinement of an AND-state.
As with OR-states, the refinement of an orthogonal component can be propagated to its inner elements. 
An AND-state can also be refined to include additional orthogonal components, which are granted the \AVirtual{} modifier by default.

\noindent\textbf{R5: Default State}\label{sec:InitialState}
Default states follow the same rules that apply to the type of state that it is.
Additional the resulting state must still be marked as default.

\noindent\textbf{R6: Final State and Fork, Split, Merge, Join Pseudo-States}\label{sec:FinalState}
These states are not directly refinable. The incoming and outgoing transitions must be preserved, though if the state is not \Locked{} additional incoming or outgoing transitions may be added in accordance with the modifiers applied. Final states do not allow for outgoing transitions.

\noindent\textbf{R7: History State}\label{sec:History}
A shallow history state is either preserved or is refined into a deep history state.
Deep history states are considered locked by default in order to preserve potential dependencies in the statechart.

\vspace{-.5\baselineskip}
\subsection{Incremental Statechart Modeling Example} \label{sec:Example}
\vspace{-\baselineskip}

\begin{figure}
  \vspace{-1.5\baselineskip}
	\centering
	\epsfig{file=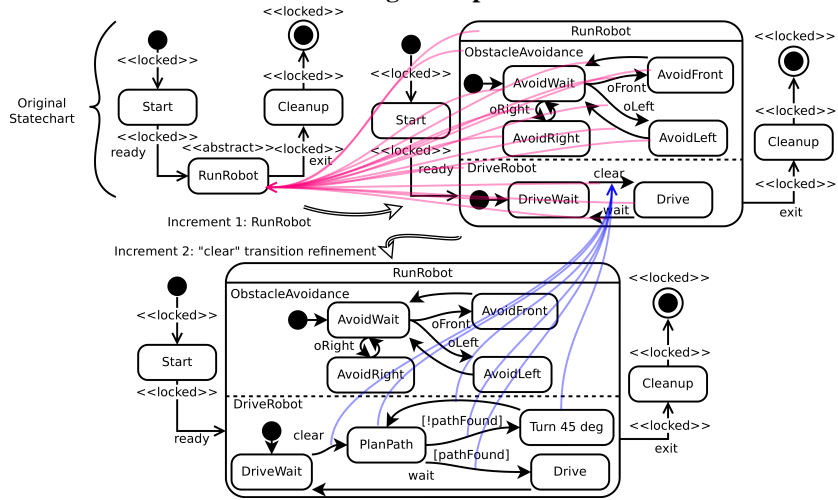, width=0.9\linewidth}
	\caption{The SimpleRobot statechart refinement example}
	\label{fig:simpleRobot}
  \vspace{-1.5\baselineskip}
\end{figure}
To illustrate the usability of the rules governing the refinement of Statecharts, consider the initial statechart in \Fig{fig:simpleRobot} that models a generic design for  robotics interface.
This structure guarantees the basic start/stop behavior of anything relying on this statechart for its functionality.
Since our rules rely on Statemate semantics, the outer transitions will always fire if able before any substate transitions.
The only refinable element is the \Model{RunRobot} abstract basic state which is to be refined in subsequent increments.
\Fig{fig:simpleRobot} also shows two increments of the initial statechart.
The pink and blue arrows represent mappings of refined elements to their original elements for each refinement.

The first refinement step applies to the \Model{RunRobot} basic state, which refines to an implicit \Standard{} AND-state with two orthogonal components.
At the beginning of this refinement step, \Model{RunRobot} would have had a \Virtual{} modifier attached to it, though this modifier was removed at the end of the refinement. This was done as an example to demonstrate that because we want certain structural guarantees, we are granting only basic freedoms to future refinements.
At this stage, the \Model{DriveRobot} orthogonal component models the movement of the robot while avoiding obstacles.
The algorithm in this increment relies on the sensor array to return a \Model{clear} event, at which point the robot will simply begin driving forward until obstacle avoidance is triggered.

A second increment models explictly the obstacle avoidance algorithm in the statechart.
In this case, the transition triggered by the event \Model{clear} is refined into multiple transitions and states.
These states are used to do very basic, na\"{\i}ve path planning starting from a waiting state.
This implementation allows for the robot to now plan a path before attempting to drive forward.
This allows us to implement a navigational algorithm inside the \Model{DriveRobot} orthogonal component, while still respecting the external entry and exit behavior.
That is, if the state \Model{DriveWait} receives the event \Model{clear}, then the token of execution will move forward until stopping at state \Model{Drive}. If state \Model{Drive} receives event \Model{wait}, then the execution token will, as expected, arrive back at state \Model{DriveWait}.
Likewise, if the \Model{exit} event was to be broadcast to this Statechart, no matter how we implement the refinement detail, the \Model{RunRobot} state will exit as expected.


We have implemented this example as a proof of concept of our rules for refinement in \ATOM{}.
The initial statechart is first modeled and annotated with modifiers.
Although this original statechart is executable, it was designed with the intention of being further refined at a later stage, specifically in the \Model{RunRobot} basic state.
This example also illustrates the need for multi-step refinements.
With the current implementation, if a transition is desired to be refined into two transitions with an AND- or OR-state in between (such as if we were refining \Fig{fig:transition} and wanted state \Model{C} to be an AND-state instead of basic) then this must be performed in two refinements.


\vspace{-.5\baselineskip}
\section{Related Work} \label{sec:relatedworks}
\vspace{-.5\baselineskip}

To the best of our knowledge, the literature has not explored the concept of Statecharts refinement in the context of incremental modeling.
The closest work has been done in $\mu-$Charts~\cite{Scholz1998,Scholz2001} using refinement calculus which we discuss later.
Instead, it is done in the context of class inheritance, where a subclass can inherit from a superclass' statechart.
We focus on three main Statecharts implementations: Rhapsody, UML State Machines, and Statemate.
Crane \etal{}\cite{Crane2007} have previously compared the semantics of these three variants, but did not discuss how they implement refinement.

\textbf{Rhapsody}~\cite{rhapsody} defines three types of modifiers on that can be annotated on a statechart: \emph{inherited}, \emph{overridden}, and \emph{regular}.

\noindent\textbf{Modifiers and Inheritance}
Elements marked as being inherited will allow for changes to the parent (original) statechart 
to propagate to the child (refined) statechart 
Elements marked as being overridden no longer accept general changes from the parent statechart other than deletions.
We do not provide this behavior.
The original-refined relationship can be severed, leaving the refined statechart in an independent state from the original.
Rhapsody additionally allows for both transitions and states to be added freely, which we only allow in states marked as \Virtual{}.

\noindent\textbf{Transitions}
In Rhapsody, events must be preserved, though the guards and any broadcast events attached to the transition can be modified arbitrarily.
We do not allow this
as this would allow for the behavior of the statechart to be altered arbitrarily. 
If a guard is desired to be removed, then either a different original statechart model should be used or it may need to be redesigned.

\noindent\textbf{States}
%
Rhapsody allows one to freely override and redefine actions on refined states.
Since modifying actions does not alter the external behavior of the original statechart, we also allow for free modification of actions.

The \textbf{Unified Modeling Language (UML)} presents three sets of inheritance rules for State Machines~\cite{OMG2001}: \emph{subtyping}, \emph{strict inheritance}, and \emph{general refinement}.
We are only concerned here with the first two sets of rules.
The focus of subtyping is to preserve the pre-/post-conditions and general behavior of a state machine, such that the refined state machine can stand in place of its parent.
This is aligned with our intentions.

\noindent\textbf{Transition subtyping}
%
UML allows for refined transitions to change their target state, which we disallow due to this breaking the external behavior of the original statechart.
As in our rules, events and sequences of events must be preserved. 
Guards are allowed to utilize disjunctions, which weakens the pre-condition required for the transition to be able to fire.
We only allow conjunctions, which strengthens the pre-condition and remains within the state space of the original statechart.
New transitions can be added as desired, which we only allow in \Virtual{} states.

\noindent\textbf{States subtyping}
%
A state can add more outgoing transitions, though it does not have to preserve its incoming transitions.
In our solution, all transitions must be at least partially preserved.
This guarantees the certain behaviors of the original statechart. 
In UML, all OR- and AND-states are granted the freedoms that we only grant in \Virtual{} states.

The focus of strict inheritance is to preserve re-use associated with inheritance.

\noindent\textbf{Transition strict inheritance}
%
The target, source, events, and sequences of events of a transition may be changed.
We do not allow for transitions to change their source or target states, 
and all events must be retained in order to preserve the external behavior of the original statechart.
As in Rhapsody, guards can be freely modified, which is not allowed in our solution.

\noindent\textbf{State strict inheritance}
States preserve their outgoing transitions, while allowing for more to be added.
In our implementation, only states inside a \Virtual{} state can have more transitions added.
As in our solution, new states can be added to OR- and AND-states within \Virtual{} states.

Strict inheritance, as with subtyping, provides much more freedom than our solution.
Our focus is to provide more predictability and clearly defined expectations of any refined statechart than the rules provided by UML.

Harel's \textbf{Statemate}~\cite{Harel1997} also addresses the concept of Statecharts inheritance, laying out several basic rules for states and transitions.
In Harel's solution, basic states can be refined into other basic states, OR-states, or AND-states.
Orthogonal components can also be added to any state, though all OR- and AND-states are granted the same freedoms as our \Virtual{} states.

Transitions can be added as desired, whereas we require refined transitions to maintain the overall pattern of connectivity of the original transition.
Statemate also allows for the target state of a transition and the guards on that transition to be changed as desired.
New broadcasts can be added, but original broadcasts must remain in order, allowing for developers to easily break the external behavior of the original statechart.

The \textbf{$\mu$-Charts formalism}~\cite{Scholz1998,Scholz2001} was introduced as a variant of Statecharts that sought to provide additional expressive power.
In these papers, Scholz defines a calculus for refining $\mu$-Charts as part of an incremental design process.

\noindent\textbf{Transitions}
Transitions can be added so long as the event associated with the new transition does not conflict with any other transition.
Our solution allows for transitions to be added only in \Virtual{} states.
This allows limited expectations to hold.
Transitions can be removed only if there is another transition present that has the same event.
Guards can be added, so long as several formal conditions hold.
We do not allow for events to be modified in our solution.

\noindent\textbf{States}
The $\mu$-Charts rules allow for the removal of initial states, which effectively amounts to the removal of an orthogonal component.
We do not allow for the removal of orthogonal components. 
Additional states can also be added, so long as they are plugged into the existing flow using new transitions, which we only allow in \Virtual{} states.

\vspace{-.5\baselineskip}
\section{Conclusion} \label{sec:conclusion}
\vspace{-.5\baselineskip}

In this paper, we presented a new refinement relation to support the control of further Statecharts refinements.
Unlike other approaches, we restrict refinements to guarantee the preservation of the external behavior of the original model.
We presented a set of refinement rules that are consistent with these constraints, while taking into account certain practical and real-world considerations, such as the introduction of refinement modifiers in the Statecharts formalism.
We implemented and tested our refinement rules on an example of incremental Statecharts development, in order to demonstrate how the original external behavior is preserved while also granting extensive developmental freedoms.
We foresee that the application of this refinement relation is not only useful for the re-use of statechart models, but can also serve to better design statechart models during the development life cycle.
For future work, we plan to review our design decisions to take into consideration alternative Statecharts semantics, since we currently only entertain the usage of Statemate semantics.

\vspace{-.5\baselineskip}
\bibliographystyle{splncs}
\bibliography{paper}

\end{document}